\documentclass{Interspeech2024}

\usepackage{enumitem}
\usepackage{tipa} 
\usepackage{xcolor}
\usepackage[absolute,overlay]{textpos} 

\interspeechcameraready

\title{What Needs to be Known in Order to Perform a Meaningful Scientific Comparison Between Animal Communications and Human Spoken Language}

\name{Roger K.}{Moore}

\address{Speech \& Hearing Research Group, Dept.\ Computer Science, University of Sheffield, UK}

\email{r.k.moore@sheffield.ac.uk}


\begin{document}

\begin{textblock*}{\textwidth}(3cm,2.5cm) 
   \emph{Proc.\ Vocal Interactivity in-and-between Humans, Animals and Robots} (VIHAR-24), Kos, Greece, 6 Sept. 2024.
\end{textblock*}

\maketitle

\begin{abstract}  
Human spoken language has long been the subject of scientific investigation, particularly with regard to the mechanisms underpinning speech production. Likewise, the study of animal communications has a substantial literature, with many studies focusing on vocalisation. More recently, there has been growing interest in comparing animal communications and human speech. However, it is proposed here that such a comparison necessitates the appraisal of a minimum set of critical phenomena: i) the number of degrees-of-freedom of the vocal apparatus, ii) the ability to control those degrees-of-freedom independently, iii) the properties of the acoustic environment in which communication takes place, iv) the perceptual salience of the generated sounds, v) the degree to which sounds are contrastive, vi) the presence/absence of compositionality, and vii) the information rate(s) of the resulting communications.
\end{abstract}

\section{Introduction}\label{sec:INTRO}

Human spoken language has been the subject of scientific investigation for a considerable period of time, particularly with regard to the mechanisms underpinning the process of speech production \cite{Fant1960,Ladefoged1962,Denes1973,Fry1979,Stevens1998}.  Likewise, the study of animal communications has a substantial history \cite{Bradbury1998,Hopp1998}, with many studies focused on the particular role of vocalisation \cite{Seyfarth2003,Fitch2006a,Seyfarth2010}.  More recently, there has been growing interest in the similarities and differences between human speech and animal communications \cite{Fitch2000,Hauser2002,Prather2013,Scott-Phillips2015b,Moore2016q,Vernes2017,TerHaar}, especially aspects of spoken language that hitherto have appeared to be unique (or at least special) in comparison with the structure of vocal communication systems observed in the rest of the animal kingdom \cite{Scott-Phillips2015,Beecher2021}.

\begin{figure}
	\centering
	\includegraphics[width=0.9\columnwidth]{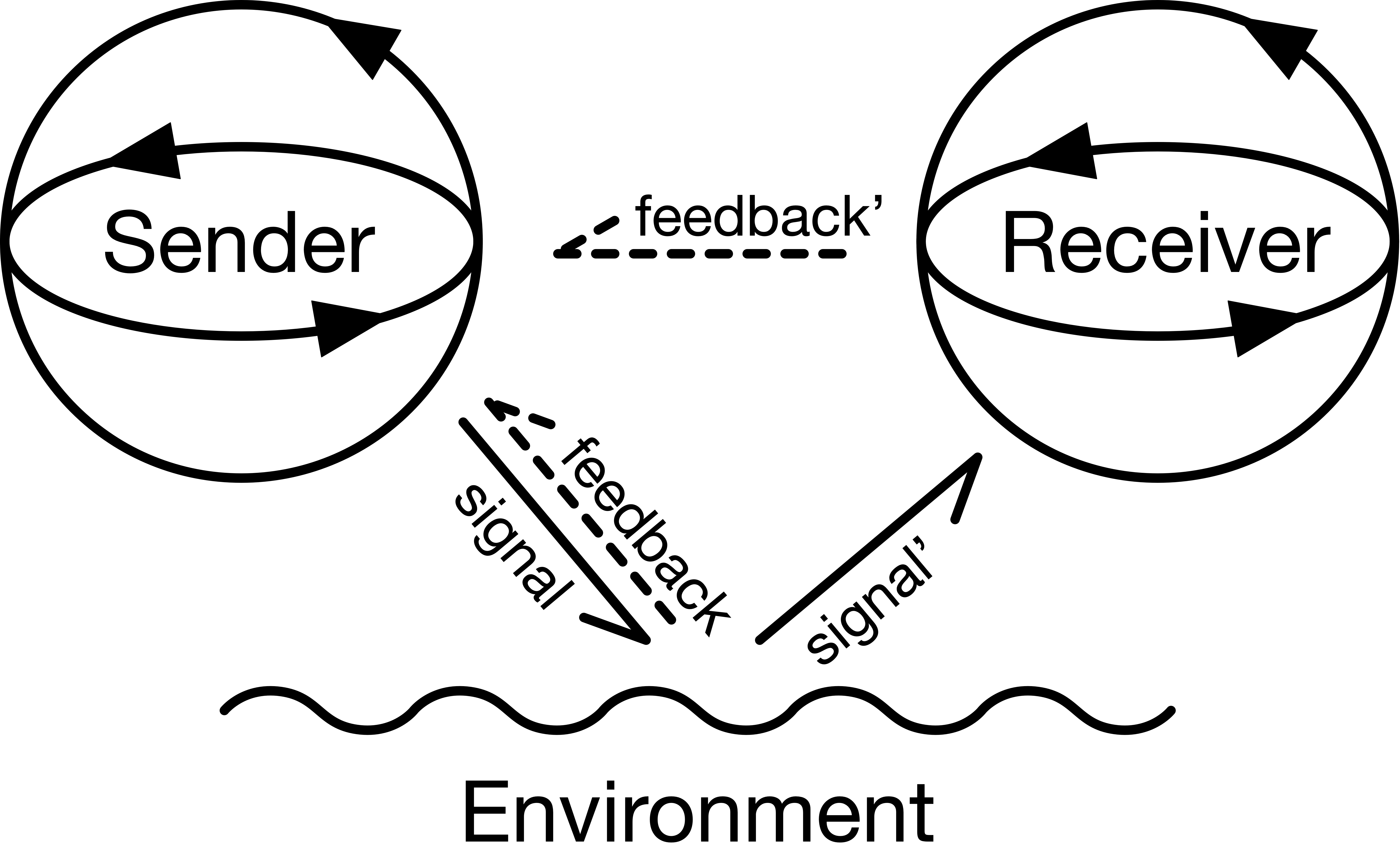}
	\caption{Illustration (using an extension of Maturana \& Varela's pictographs \cite{Maturana1987,Moore2016s}) of communication between sender and receiver `cognitive unities' (human beings or animals) via a conditioning environmental context.}
	\label{fig:PIC}
\end{figure}

Of course, there are many ways in which human and animal behaviour may be compared, and the approach taken very much depends on the interests of the individual researchers and their home fields of study.  However, comparing animal communications and human speech requires a particularly careful appraisal of a number of phenomena relating to the production, transmission and reception of communicative signals, while simultaneously taking into account the social and pragmatic contexts within which such interactions take place (see Fig.~\ref{fig:PIC}).  This is not easy to do -- especially for animals!  Nevertheless, an important methodological step is to identify a minimum set -- a `checklist' -- of critical phenomena that need to be characterised in order to perform a meaningful scientific comparison between animal communications and human spoken language.  This paper puts forward such a checklist.

\section{What we need to Know}\label{sec:WHAT}

It is proposed that,  in order to perform a scientific comparison between animal communications and human spoken language, the minimum that needs to be known is \ldots
\begin{enumerate}[leftmargin=20pt,label=\roman*]
	\item the number of degrees-of-freedom of the vocal apparatus,
	\item the ability to control those degrees-of-freedom independently,
	\item the properties of the acoustic environment in which communication takes place,
	\item the perceptual salience of the generated sounds,
	\item the degree to which sounds are contrastive,
	\item the presence/absence of compositionality, and
	\item the information rate(s) of the resulting communications.
\end{enumerate}

\subsection{Degrees-of-freedom of the vocal apparatus}\label{sec:DOF}

The term `degrees-of-freedom' (DoF) was originally established in the field of statistics to characterise the number of \emph{independent} pieces of information that contribute to estimating the value of a parameter.  More recently, DoF has been used in robotics to refer to the number of independent articulators (which is usually directly related to the number of actuators or motors).  The concept captures the \emph{dimensionality} of the space of possible physical movements, and is thus highly relevant to characterising their potential use in signalling/communications.

In principle, the number of DoFs of the vocal apparatus is derivable from measurements of the anatomy and the physics of sound production.  However, unlike an artificial device such as a robot, a natural living system possesses a very large (effectively, infinite) number of degrees-of-freedom.  In this case, statistical analysis of the movements -- e.g.\ by `principal components analysis' (PCA) -- can provide an estimate of the underlying dimensionality (i.e.\ how the anatomy is used in practice) and thereby inform the structure of an appropriate mathematical model.

A common model of animal vocalisation (especially human speech) is the `source-filter' model \cite{Fant1960}, in which the excitation provided by a sound source (such as the vibration of vocal folds in a larynx, or the actions of a syrinx) is modelled separately from the resonant properties of an acoustic-tube approximation of the vocal tract.  Modelling the latter using `linear prediction analysis' (LPA) \cite{Atal1971}\footnote{It may be interesting to note in passing that LPA allows the quality of the information present in the source and filter paths to be modified independently, and thus could facilitate a novel means for investigating animal communications (particularly with regard to the topics addressed in Sections~\ref{sec:PERC}  and \ref{sec:CNTRST} ).}, `formant' resonators \cite{Klatt1980,Holmes1983} or an `acoustic-waveguide' \cite{Mullen2003,Story2005} can provide accurate simulations of human speech, and can be extended to many other animals (especially non-cetacean mammals) \cite{Moore2016p,Anikin2019}.

However, a malleable sound source or a deformable tube has many potential DoFs.  Hence, a crucial factor for communication is the degree to which they are under active \emph{control}.

\subsection{Control}\label{sec:CNTRL}

`Control Theory' is an established discipline in the field of engineering \cite{DiStefanoIII1990}, and `Perceptual Control Theory' (PCT) is the application of control theory to modelling the behaviour of living systems \cite{Powers1973,Mansell2015}.  Derived from `cybernetics' \cite{Wiener1948}, a key notion is the use of \emph{feedback} to regulate an intended control action.  In particular, \emph{closed-loop} control using negative feedback provides a simple yet powerful mechanism for stabilising behaviour in the face of unknown disturbances (just as a thermostat is able to maintain the ambient temperature in a room despite doors and windows being constantly opened and closed).

Of particular interest here are: i) the \emph{number} of DoFs under active control (i.e.\ the ability of an animal or human being to control those degrees-of-freedom independently), and ii) the \emph{quality} of control for each DoF in terms of their temporal and positional precision as well as their resistance to disturbance.  In other words, for communications it is not enough to know what DoFs are being controlled; it is also necessary to know how feasible it is for a \emph{sender} to achieve particular motor targets in a reliable and timely manner, and whether such targets can be maintained in the presence of disturbances\footnote{Note that this may be categorised as \emph{sender-oriented} control.} \cite{MacDonald2011,Kim2014,Brainard2000}.

\subsection{Acoustic environment}\label{sec:ENV}

Once a communicative signal has been generated by a sender, it has to propagate through the environment to a \emph{receiver} (as illustrated in Fig.~\ref{fig:PIC}).  Clearly, the acoustic characteristics of the environment will impact on how (or whether) the signal is perceived by the receiver.  However, the environment may also have an impact on the sender, either through long-term (phylogenetic) adaptation \cite{Endler1992,Forrest1993} or, more interestingly, via short-term (feedback) control \cite{Lombard1911,Brumm2006}, in which case the environment may be viewed as a potential disturbance\footnote{Note that this may be categorised as \emph{receiver-oriented} control.}.  Hence, determining the level of dependency between emitted signals and the communicative environment is a crucial piece of information in the context of comparing animal communications with human speech.

\subsection{Perceptual salience}\label{sec:PERC}

Once a communicative acoustic signal arrives at the ears of a listener, it is not only important that it is heard (above any ambient noise/interference), but also that any crucial distinctions are actually perceived as different.  In other words, there is no value in a sender crafting subtle differences between signals if a receiver is unable to discriminate between them.  Hence, it is necessary to understand the psychophysics of listeners' perceptual acuity, for example by characterising their ability to detect `just-noticeable-differences' (JNDs).  However, although measuring JNDs for human listeners is a well-established procedure \cite{Cheatham1950}, it is considerably more difficult to perform on animals \cite{Ndez-Juricic2016}.

Of course, senders are usually also receivers, which means that they are (in principle) able to assess the salience of their own communicative emissions.  However, this strategy is only valid for communication between conspecifics; communication between different animals, between humans and animals or even between humans and artificial agents will inevitably be limited by any mismatch in perceptual capabilities \cite{Moore2016x,Huang2022}.

\subsection{Contrastive signalling}\label{sec:CNTRST}

While it is important for a sender to create perceptible distinctions between each item in their inventory of signals, for a living system there is another factor at play -- `energetics' -- that is, the degree of physical and/or neurological \emph{effort} involved in the process.  This means that, in principle, by increasing the level of effort, it is not only possible for a sender to optimise perceptual salience by making signals louder, but also by making them \emph{clearer} (i.e.\ more distinct from one another).  Of course, the active management of effort is dependent on a sender's \emph{motivation} to do so, and thus linked to their situational context, for example there may be a degree of urgency associated with the communications.

In human speech, the ability to vary the clarity of a signal along a continuum from \emph{hypo}-articulation (mumbling) to \emph{hyper}-articulation (clear speech) is described by `H\&H Theory' \cite{Lindblom1990} which posits that sound production is actively managed using a closed-loop control process (as already discussed in Section~\ref{sec:CNTRL} above).  Not only does this facilitate the dynamic adjustment of speech intelligibility in the face of arbitrary environmental disturbances such as noise or reverberation (see Fig.~\ref{fig:HAWK}), but it has also led to a system of communication in which sounds are used in a \emph{contrastive} manner, i.e.\ to distinguish one meaning from another.  This is manifest as the `phonemic' structure of human speech whereby acoustically distinct speech sounds are \emph{only} perceived as different (by native listeners) \emph{if} they signal the difference between one word and another (in their language) \cite{Jones1973}.  Crucially, acoustically distinct speech sounds are perceived as the \emph{same} if they do \emph{not} signal the difference between one word and another.  That is, the sounds listeners perceive - the `phonemes' - are conditioned on the \emph{meaning} of an utterance, not on a fixed set of acoustic properties.  Unfortunately, this dual language-independent `phonetic' (\emph{physiophonic}) and language-dependent `phonemic' (\emph{psychophonic}) nature of speech is not always appreciated, with the consequence that the term `phoneme' is often misused \cite{moore19_interspeech}.

\begin{figure}[h]
	\center{\noindent\emph{``I! ... DO! ... NOT! ... KNOW!'' \\
	``I do not know'' \\
	``I don't know'' \\
	``I dunno'' \\
	``dunno''} \\
	\textipa{[\~@\~@\~@]}}
	\caption{An illustration of contrastive behaviour in everyday human conversation.  On hearing a verbal enquiry from a family member as to the whereabouts of some mislaid object, the listener might reply with any of the utterances shown (all of which would be perceived as ``I do not know'') \cite{Hawkins2003a}.  The particular utterance emitted would depend on the communicative context; the shouts would be necessary in a noisy environment, the nasal grunts would be sufficient in a quiet environment.}
	\label{fig:HAWK}
\end{figure}

While these contrastive behaviours are an emergent consequence of the active management of energetic constraints, and thus would seem to reflect a general principle that could apply to \emph{all} communicative behaviour, it has yet to be shown that this is the case -- especially for animal communications -- although some studies have addressed the issue \cite{Bowling2015,Engesser2015,Lachmann2001}.

\subsection{Compositionality}\label{sec:COMP}

One of the distinguishing features of human spoken language is the `particulate' nature of the sound system \cite{Abler1989}.  That is, just as chemical elements do not blend together, but combine to form structures with quite different properties to their constituent parts, so sounds may be used in different combinations to signify completely unrelated \emph{meanings}.  For example, a vocal production systems with $d$ independent DoFs each capable of producing $s$ distinct signals can generate up to $s^d$ different sounds, which means that a sequence of $n$ sounds can support up to $(s^d)^n$ different meanings.  As Alexander von Humboldt observed nearly two-hundred years ago: ``\emph{language makes infinite use of finite media}'' \cite{VonHumboldt1836}.

Clearly, exploiting combinatorics through the efficient \emph{re-use} of sub-structures is an effective means for expanding the expressive power of a communications system.  It also provides a means for composing new meanings out of old meanings (which is, indeed, a blending operation).  Whether meaningful sequences are formed by combination or by composition, these processes give rise to repetitive sound patterns.  Hence, there is interest in algorithms for detecting such repetition in human speech \cite{Stouten2008,Aimetti2010} and for determining whether such repetition occurs in animal communications \cite{Kershenbaum2015,Kershenbaum2016,Sharma2024}.

\subsection{Information rates}\label{sec:INFO}

A prime concern in speech-based interaction is \emph{what} people say, and considerable research resources have been devoted to characterising such behaviour at the traditional acoustic, phonetic, phonological, morphological, lexical, syntactic and semantic levels of description.  Such studies involve a multitude of approaches to characterising the complexity of spoken language \cite{Gibbon1997}, but `information theory' \cite{Shannon1949a,Shannon1951} provides a particularly powerful paradigm for a single unified approach to \emph{quantitative} measurement.  For example, Coup{\'{e}} et al.\ have shown that \emph{all} human languages have an information rate of $\sim$39 bits/sec at the phonetic level \cite{Coupe2019}, and Bergey \& DeDeo estimate that the information density at the lexical level is $\sim$13 bits/sec \cite{Bergey2024}.

Similar principles have been applied to animal communications, e.g.\ entropic values have estimated for bottlenose dolphin whistles and squirrel monkey chucks \cite{McCowan2002}.  Likewise, considerable effort has been devoted to understanding the appropriate methodology \cite{Kershenbaum2014a}.  In the context of this paper, all of the considerations discussed in Sections~\ref{sec:DOF} to \ref{sec:COMP} (especially, DoFs and JNDs) could be characterised using an information theoretic approach, thereby providing a unified method for comparison across different species.

In reality, human spoken language and animal communications is unlikely to be a fixed code with a \emph{constant} information rate.  The information that is communicated is inevitably going to be conditioned on critical causal variables \cite{Moore2023c} such as \ldots
\begin{itemize}[leftmargin=20pt]
	\item the situated and embodied context (i.e.\ \emph{pragmatics}),
	\item the temporal evolution of events (i.e.\ \emph{synchronics}), and
	\item the level of effort that participants are prepared to devote to communicative behaviour (i.e.\ \emph{energetics}).
\end{itemize}

In other words, a key question in communication is not just \emph{what} is communicated, but \emph{why}, \emph{when} and \emph{how} it is communicated -- and these factors will be reflected in a local variation in information rate, e.g.\ on encountering local minima in cooperative interaction \cite{Moore2023}, or as a function of cognitive load in unstructured human conversation \cite{Bergey2024}).

\section{Summary and Conclusion}\label{sec:CONC}

This paper has proposed that, in order to perform a meaningful scientific comparison between animal communications and human speech, the minimum that needs to be known is \ldots
\begin{enumerate}[leftmargin=20pt,label=\roman*]
	\item the number of degrees-of-freedom of the vocal apparatus,
	\item the ability to control those degrees-of-freedom independently,
	\item the properties of the acoustic environment in which communication takes place,
	\item the perceptual salience of the generated sounds,
	\item the degree to which sounds are contrastive,
	\item the presence/absence of compositionality, and
	\item the information rate(s) of the resulting communications.
\end{enumerate}

The claim that this list constitutes \emph{the} minimal set of phenomena is a strong one, and it implies that these criteria are somehow unique.  This claim is justified on the basis that each item in the list represents a crucial link in the communications chain between sender and receiver (as illustrated in Fig.~\ref{fig:PIC}).  Failing to characterise one or more of these phenomena would render an overall comparison lacking in important details.

Having said that, as has been made clear in Section~\ref{sec:WHAT}, it is not the case that these topics have hitherto been ignored.  Quite the contrary, it is acknowledged that many of these areas have already been the subject of extensive investigation.  However, it is suggested here that they have often been pursued somewhat independently.  Hence, it is posited that there is value in reiterating the dependencies that exist within a communications chain, especially with regard to highlighting \textbf{closed-loop control} as a ubiquitous mechanism for regulating behaviour and \textbf{information theory} as a universal means for quantifying the relevant outcomes at all points along the chain.

It should also go without saying that none of these aspects of communications are particularly easy to characterise, particularly in animals.  However, the claim being made here is that without knowing the answers to these questions, it will be next-to-impossible to draw meaningful comparisons across species.  It is therefore hoped that this approach will stimulate productive interdisciplinary discussion.

Finally, although this paper has focused on \emph{vocal} communications, the same principles apply to \emph{multimodal} communications, i.e.\ gestures, body pose, facial expressions, eye gaze, etc.   The principles expounded here would then encompass i) the distribution of information across the available modalities, and ii) the dynamics of shifting the emphasis from one modality to another as a function of the changing communicative context.

\section{Acknowledgements}

The issues addressed in this paper were inspired by an invitation to review a paper on the topic of animal vocalisation and its potential relation to human speech.  I therefore wish to thank and acknowledge -- albeit anonymously -- the authors of the said manuscript for stimulating the particular train of thought that has been outlined here.


\bibliographystyle{IEEEtran}
\bibliography{RKM-VIHAR24}

\end{document}